\documentclass[english,french]{article}
\usepackage[T1]{fontenc}
\usepackage[latin9]{inputenc}
\usepackage{geometry}
\usepackage{array}
\usepackage{float}
\usepackage{units}
\usepackage{textcomp}
\usepackage{multirow}
\usepackage{graphicx}
\usepackage[numbers]{natbib}

\makeatletter

\providecommand{\tabularnewline}{\\}

\newcommand{\lyxaddress}[1]{
\par {\raggedright #1
\vspace{1.4em}
\noindent\par}
}

\@ifundefined{date}{}{\date{}}
\makeatother

\usepackage{babel}
\begin{document}

\title{Time-scale characteristics of Kasaï river hydrological regime variability
for 1940-1999}

\author{Jean-Marie Tshitenge Mbuebue\\
Albert Mbata Muliwavyo\\
Vincent Lukanda Mwamba\\
Edmond Phuku Phuati\\
Albert Kazadi Mukenga Bantu\\
Franck Tondozi Keto }
\maketitle

\lyxaddress{\begin{center}
Department of Physics, University of Kinshasa, B.P. 190 Kinshasa
XI, DR Congo
\par\end{center}}
\begin{abstract}
The present study was undertaken with the aim of contributing to the
characterization of the nonstationary variability of the hydrological
regime of the Kasaï River using the wavelet analysis for 1940-1999.
The rainfalls and discharge over Kasaï Basin have marked fluctuations
with a perceptible downward trend and some shift around 1950, 1960,
1970, 1983 and 1994. The results show that rainfalls over Kasaï basin
and the discharge at Ilebo station patterns exhibit a strong annual
oscillation and some intermittent oscillations in 2-8 years (1950-1975,
1983-1995) and 8-16 years (1970-1999) time scales. The wavelet coherence
analysis reveals a weak possible connection between hydrological variables
(rainfalls, discharge) and climate indices relative to sea surface
temperature and atmospheric circulation over Atlantic tropical, Indian
and Pacific Oceans (coherence less than 0.55).

\end{abstract}

\section{INTRODUCTION}

The Kasaï river is the chief southern tributary of the Congo River,
into which, at Kwamouth, Congo (Kinshasa), 200 km above Malebo (Stanley)
Pool, it empties a volume approaching one-fifth that of the main stream.
This stretch is interrupted by several spectacular rapids and waterfalls,
the river flowing in deeply trenched valleys at elevations of about
600\textendash 900 m. The Kasaï River eventually crosses a further
series of rapids and falls, broadening and deepening to make navigation
possible.

For effective and sustainable management of water resources in the
Kasaï basin the knowledge of their availability and their variability
is needed. It is also an important indication for modeling and exploration
of the evolution of water resources of this region. In order to improve
the understanding of the influence of various factors on the evolution
of the hydrological response of the Kasaï, and more broadly on the
development of water resources, from the regional to a global scale,
Kasaï watershed has been subject of few studies undertaken by De Baker
\citep{key-1}, Devroey \citep{key-2} who provided a lot of information\textquoteright s
about Kasaï basin. Ntombi et al. \citep{key-3} and Kisangala et al.
\citep{key-4} have focused on statistical analyzes based on existing
data to establish an understanding of hydrology of the basin.

In the context of climate change, the determination of preferential
scales between climate variability and hydrological variability of
Kasaï watershed is needed. The overall variability of short-term climate
is generally associated with phases of oceanic and atmospheric phenomena
coupled with El Niño (ENSO) and the North Atlantic Oscillation (NAO).
While El Niño Southern Oscillation (ENSO) affects weather and climate
variability around the world, the North Atlantic Oscillation (NAO)
is the climate dominant mode in the region of the North Atlantic.
These oscillations have been used in several studies to develop accurate
models able to predict climate variability.The determination of the
climate change impact on hydrological systems and their water resources
constitutes a major issue of the 21st century for which scientists
must answer.

The main objectives are to:
\begin{itemize}
\item identify and quantify the main modes of variability of the climate
and hydrological response of Kasaï river basin; 
\item characterize the relationships between discharge at Ilebo\textquoteright s
station, gridded rainfall over Kasaï basin signals and some climate
indices. 
\end{itemize}

\section{STUDY AREA AND DATA}

\begin{figure}[H]
\begin{centering}
\includegraphics[scale=0.65]{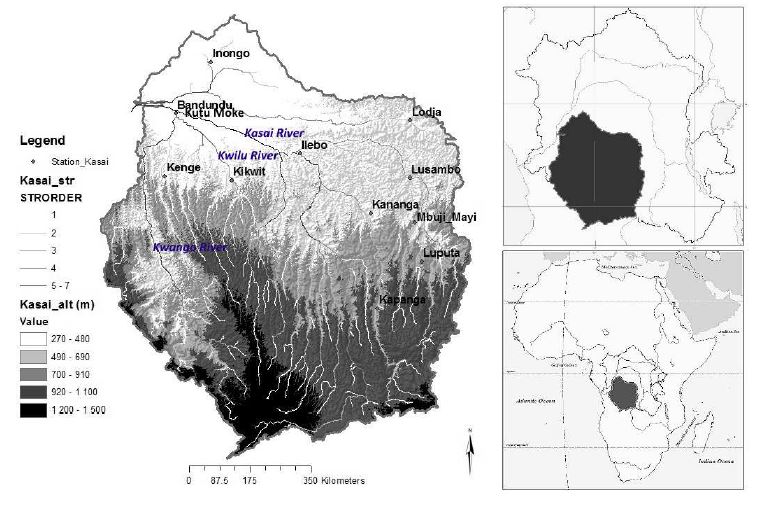}
\par\end{centering}
\caption{\label{fig:Relief-and-stream}Relief and stream order map of Kasaï
watershed}

\end{figure}

Kasaï basin covers an area of 904,000 square kilometers, between longitudes
15.75 \textdegree{} and 24.7 \textdegree{} east and between 0.75 \textdegree{}
South and 12.25 \textdegree{} South latitude (Fig. \ref{fig:Relief-and-stream}).
72.4\% of this basin is in DR Congo and the remaining 27.6\%, representing
the Southwest part is located in Angola. 

From orographic point of view (Fig. \ref{fig:Relief-and-stream}),
one can identified a depression, sprinkled with few isolated massifs,
at the North of the basin around the 4th parallel which is between
300 and 500 meters of altitude. The trays exceeding the altitude of
1000 meters are located on the western edge of the basin between Wamba
and Loange, to the south, as well as in the Southeast part. The rapids
that cut most of the major tributaries of the basin are located between
the plateau at 500 meters of altitude and the central depression.

The longest river in the southern Congo River basin system, the Kasaï
River, measures 2,153 km from its source on the eastern slope of the
Bie Plateau in Angola by 12 \textdegree{} South and 19 \textdegree{}
East longitude Greenwich, near the plateau 1500 meters above sea level
where the Zambezi River also takes its origins \citep{key-2} to Kwamouth.
The Kasaï River is the largest tributary of the Congo River. It swells
waters of many tributaries whose some of them measure themselves more
than 1,000 kilometers.

Its headstream runs east for 250 miles (402 km), then turns north
for nearly 300 miles (483 km) to form the frontier between Angola
and Congo. Traffic is especially heavy on the 510-mile- (820-kilometre-)
long waterway from Kinshasa to Ilebo. Above Ilebo, navigation is impeded
by sandbanks, but shallow-draft boats can reach Djokupunda. There
the Mai- Munene Falls require goods to be carried 40 miles (64 km)
by rail to the next navigable reach, from Mukumbi to Mai- Munene.
In its lower reaches, the river enters the equatorial rainforest.
Below its confluence with the Kwango, it forms Wissmann Pool and then
receives the Fimi-Lukenie, bringing the waters of Lake Mai-Ndombe.
Thereafter, it is known as the Kwa.

The climate is equatorial in the northern part of Kasaï basin. It
is characterized by little difference in seasons, with high temperature
and humidity virtually constant. The rainfall distribution throughout
the year is on average around 1500 millimeters. 

\begin{figure}[H]
\begin{centering}
\includegraphics[scale=0.9]{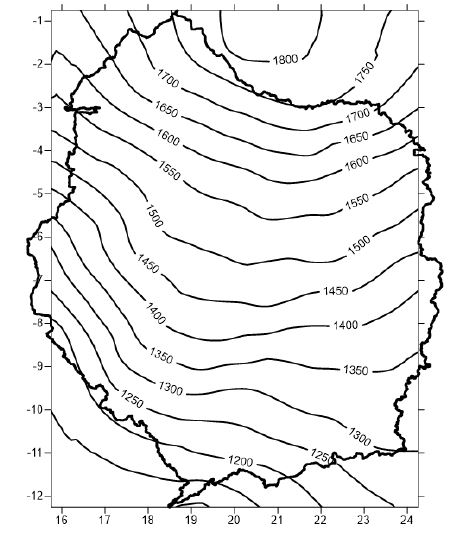}
\par\end{centering}
\caption{\label{fig:Mean-annual-rainfall}Mean annual rainfall map of the Kasaï
basin (1940-1999). Data: SIEREM.v1 at a $0.5{^\circ}\times0.5{^\circ}$
resolution. Isohyets in mm.}

\end{figure}

\begin{figure}[H]
\begin{centering}
\includegraphics[scale=0.9]{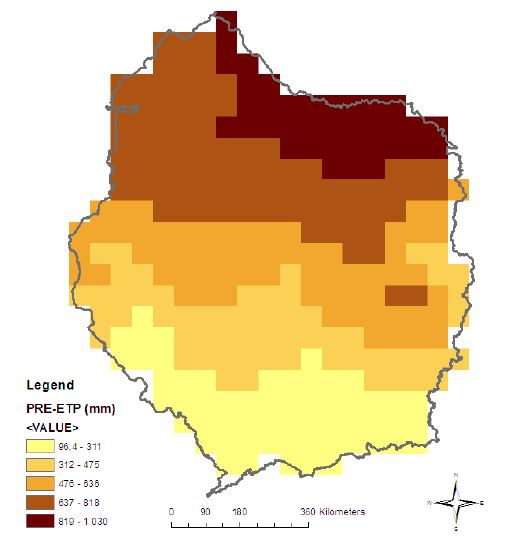}
\par\end{centering}
\caption{\label{fig:Rainfalls-minus-Evapotranspirati}Rainfalls minus Evapotranspiration
(PRE-ETP) map over Kasaï watershed the Kasaï basin (1940-1999). Data:
Climate Research Unit CRU CL 3.10 climatology at a 0.5\textdegree{}
resolution (New et al., 2002). Isohyets in mm.}

\end{figure}

In the South, there is the tropical climate of the Sudan type with
two maxima of rainfall and temperature; the dry season becomes longer
as one progresses from the north towards the south in the basin. The
relief of the southern region of Kasaï basin causes a slight temperature
drop. 

Variations in the length of the rainy season depending on the latitude
are clearly demonstrated by the graph in Fig. \ref{fig:Typology-of-rainfall}.
This basin had 38 weather stations in the colonial era. The mean maximum
and minimum absolute temperatures are respectively 32 \textdegree /27
\textdegree C and 21 \textdegree /13 \textdegree C in the north/south
of Kasaï basin (Fig. \ref{fig:a)-Minimum,-b)}).

\begin{figure}[H]
\begin{centering}
\includegraphics[scale=0.5]{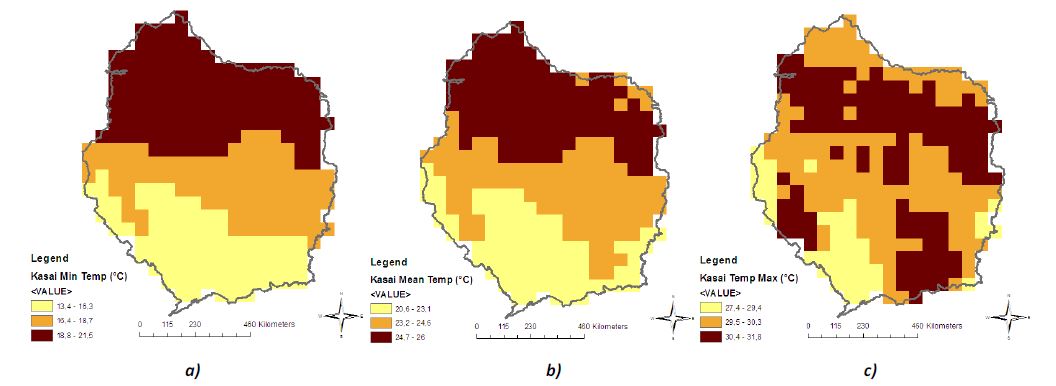}
\par\end{centering}
\caption{\label{fig:a)-Minimum,-b)}a) Minimum, b) Mean and c) maximum temperature
over Kasaï watershed the Kasaï basin (1940-1999). Data: Climate Research
Unit CRU CL 3.10 climatology at a 0.5\textdegree{} resolution (New
et al., 2002). Isohyets in mm.}

\end{figure}

The great equatorial forest covers just about any depression of the
central basin until about the 5th parallel; it pushes its ramifications
as wide wooded galleries, going up through the valleys of Kwilu, Kasaï
and Lulua until the 7th parallel.

\subsection*{Climate data}

Gridded rainfall (1940-1999) were extracted from the African gridded
data set from the SIEREM database \citep{key-5} provided by the HydroSciences
Montpellier Laboratory on the basis of several data sources, including
the former ORSTOM (now IRD \textendash{} Institut de Recherche pour
le Développement) rainfall database \citep{key-6}. This data set
SIEREM.v1 was constructed from SIEREM database by kriging interpolation.
SIEREM is an environmental information system for water resources.
Raw historical climate datasets are not appropriate for statistical
analysis because of inhomogeneity and leakage while gridded datasets
have undergone rigorous quality control and have been homogenized
\citep{key-7,key-8}.

Climate indices used are listed in table \ref{tab:-Data-sources}.

\begin{table}[H]

\caption{ \label{tab:-Data-sources}Data sources used for monthly climate indices }

\centering{}%
\begin{tabular}{|c|c|c|c|}
\hline 
{\footnotesize{}Climate indices} & {\footnotesize{}Data} & {\footnotesize{}Periods} & {\footnotesize{}Source}\tabularnewline
\hline 
\hline 
{\scriptsize{}NAO} & {\scriptsize{}North Atlantic Oscillation \citep{key-9}} & {\scriptsize{}1950-2013} & {\scriptsize{}\textquotedblleft Earth System Research Laboratory (ESRL)}\tabularnewline
\hline 
{\scriptsize{}Nino 3.4} & {\scriptsize{}East Central Tropical Pacific SST Index \citep{key-10}} & {\scriptsize{}1951-2013} & {\scriptsize{}of the National Oceanic and}\tabularnewline
\hline 
{\scriptsize{}TNA} & {\scriptsize{}Tropical Northern Atlantic Index \citep{key-11}} & {\scriptsize{}1948-2013} & {\scriptsize{}Atmospheric Administration (NOAA)\textquotedblright{}}\tabularnewline
\hline 
{\scriptsize{}TSA} & {\scriptsize{}Tropical Southern Atlantic Index \citep{key-11}} & {\scriptsize{}1948-2013} & {\scriptsize{}web page http://www.cpc.ncep.noaa.gov/data/ind ices/}\tabularnewline
\hline 
{\scriptsize{}Solar} & {\scriptsize{}Solar Flux \citep{key-12}} & {\scriptsize{}1948-2014} & \tabularnewline
\hline 
{\scriptsize{}DMI} & {\scriptsize{}Dipole Mode Index \citep{key-13}} & {\scriptsize{}1870-2014} & {\scriptsize{}The Frontier Research Center for Global Change}\tabularnewline
\hline 
 &  &  & {\scriptsize{} (JAMSTEC, Japan)\textquotedblright{} web page}\tabularnewline
\hline 
 &  &  & {\scriptsize{}(http://www.jamstec.go.jp/frcgc/resear ch/d}\tabularnewline
\hline 
 &  &  & {\scriptsize{}1/iod/HTML/Dipole\%20Mode\%20In dex.html)}\tabularnewline
\hline 
\end{tabular}
\end{table}

\section{METHODS}

Characterization of the hydrological variability (discharge), the
rainfall fluctuations and their comparison with the climate indices
are performed by using the signal processing methods (multitaper,
wavelet analysis,\dots ). 

Climate variability refers to the climatic parameter of a region varying
from its long-term mean. Every year in a specific time period, the
climate of a location is different. Some years have below average
rainfall, others have average or above average rainfall.

\begin{figure}[H]
\begin{centering}
\includegraphics[scale=0.7]{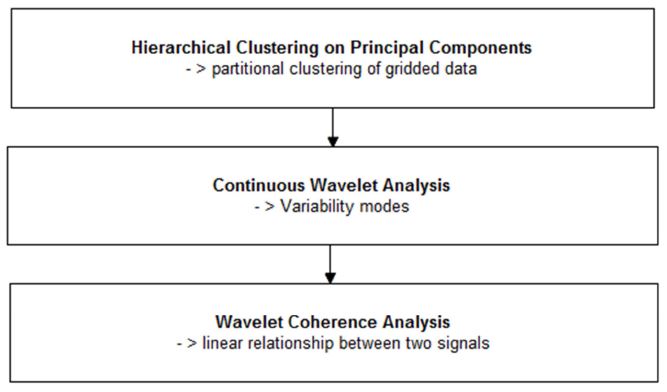}
\par\end{centering}
\caption{\label{fig:Diagram-of-the}Diagram of the overall approach of the
study}

\end{figure}

\subsection{HIERARCHICAL CLUSTERING ON PRINCIPAL COMPONENTS}

Combining principal component methods, hierarchical clustering and
partitional clustering, help to better visualize data and highlight
the main features of the data set \citep{key-14,key-15}.

\subsection{THE SEGMENTATION METHOD OF HILBERT}

The time series segmentation procedure in this case gives several
changes of average. By using a specific algorithm, one or more shift
dates (if any) which separate adjacent segments whose averages have
a significant difference under the Scheffe test are selected \citep{key-16}.

\subsection{WAVELET TRANSFORM}

Recently the wavelet transform has gained a lot of popularity in the
field of signal processing. This is due to its capability of providing
both time and frequency information simultaneously, hence giving a
time-frequency representation of the signal. The traditional Fourier
transform can only provide spectral information about a signal. Moreover,
the Fourier method only works for stationary signals. In many real
world applications, the signals are non-stationary. One solution for
processing non-stationary signals is the wavelet transform. The wavelet
analysis allows the passage of a representation of a signal to another
as Fourier analysis, with a different time-frequency resolution. The
wavelet transform $W_{n}(s)$ of a discrete time series xn is defined
as the convolution of xn with a scaled and translated version of a
function l/JO called « mother wavelet » \citep{key-17}: 

\begin{equation}
W_{n}\left(s\right)=\sqrt{\frac{\delta t}{s}}\sum_{n'=0}^{N-1}x_{n'}\psi_{0}\left[\frac{\left(n'-n\right)\delta t}{s}\right]
\end{equation}

where $\delta t$ and $s$ are respectively the time step and the
scale. 

In this paper, Morlet wavelet is used as « mother wavelet » and consists
of a plane wave modulated by a Gaussian: 

\begin{equation}
\psi_{0}\left(\eta\right)=\pi^{-\frac{1}{4}}e^{i\omega_{0}\eta}e^{-\frac{\eta^{2}}{2}}
\end{equation}

and $\omega_{0}$ is taken to be 6 \citep{key-18}.

\subsection{GLOBAL WAVELET SPECTRUM (GWS) OR TIME AVERAGE OF LOCAL WAVELET SPECTRA}

The Global Wavelet Spectrum (GWS) is used to estimate changes of wavelet
power with frequency \citep{key-20}. It provides an unbiased and
consistent estimation of the true power spectrum of a time series.
:

It is calculated by: 

\begin{equation}
\overline{W}^{2}\left(s\right)=\frac{1}{N}\sum_{n=0}^{N-1}\left|W_{n}\left(s\right)\right|^{2}
\end{equation}

\subsection{SCALE-AVERAGED WAVELET POWER (SAWP)}

The Scale-Averaged Wavelet Power (SAWP), which is defined as the weighted
sum of the wavelet power, is used to examine fluctuations in power
over a range of a scales (a band). It is given by: 

\begin{equation}
\overline{W}_{n}^{2}=\frac{\delta_{j}\delta_{t}}{C_{\delta}}\sum_{j=j_{1}}^{j_{2}}\frac{\left|W_{n}\left(s_{j}\right)\right|^{2}}{s_{j}}
\end{equation}

where $C_{\delta}=0.776$ for the Morlet wavelet, $\delta_{j}$ is
the factor of scale averaging, $\delta_{t}$ is the sampling period,
$j_{l}$ and $j_{2}$ are the scales over which SAWP is computed. 

The wavelet \textendash filtered reconstructions of a determined oscillation
band-passed, $j_{l}$ and $j_{2}$, is given by: 

\begin{equation}
X_{n}=\frac{\delta_{j}\delta_{t}^{\frac{1}{2}}}{C_{\delta}\psi_{0}\left(0\right)}\sum_{j=j_{1}}^{j_{2}}\frac{R\left[W_{n}\left(s_{j}\right)\right]}{s_{j}^{\frac{1}{2}}}
\end{equation}

where the factor $\psi_{0}\left(0\right)$ is used for removes the
energy scaling. 

\subsection{WAVELET COHERENCE ANALYSIS }

The wavelet coherence, $R_{n}^{2}(s)$, is defined as the square of
the cross-spectrum normalized by the individual power spectra \citep{key-19}. 

\begin{equation}
R_{n}^{2}\left(s\right)=\frac{\left|s\left(s^{-1}W_{n}^{XY}\left(s\right)\right)\right|^{2}}{s\left(s^{-1}\left|s^{-1}W_{n}^{X}\left(s\right)\right|^{2}\right)s\left(s^{-1}\left|s^{-1}W_{n}^{Y}\left(s\right)\right|^{2}\right)}
\end{equation}

where $R_{n}^{2}(s)\in\left[0,1\right]$, $W_{n}^{XY}\left(s\right)=W_{n}^{X}\left(s\right)\times W_{n}^{Y}(s)$
is the cross-wavelet, $n$ represents the time index, $s$ the scale
of $Y$ wavelet transform, $W_{n}^{X}\left(s\right)$ and $W_{n}^{Y}\left(s\right)$
are the wavelet transform of $X$ and $Y$ and $S$ denotes the smoothing
operator for $W_{n}^{XY}\left(s\right)$. 

The wavelet phase difference or relative phase between two time series
$X$ and $Y$ is given as: 

\begin{equation}
\phi_{n}=\arctan\nicefrac{\mathcal{J}\left(s^{-1}W_{n}^{XY}\left(s\right)\right)}{\mathcal{R}\left(s^{-1}W_{n}^{XY}\left(s\right)\right)}
\end{equation}

where $\mathcal{J}\left\{ .\right\} $ and $\mathcal{R}\left\{ .\right\} $
are the imaginary and real parts of the wavelet spectra. Wavelet coherence
lies between 0 and 1. This reflects the time\textendash scale variability
of the linear relationship between two signals. A near-zero coherence
indicates no linear relationship between the signals. Coherence close
to 1 indicates a linear relationship between both signals.

\section{RESULTS}

Hierarchical Clustering on Principal Components Kasaï basin rainfalls
shows that the Kasaï basin can be divided into three areas as in fig.
\ref{fig:Map-of-the}. 

\begin{figure}[H]
\begin{centering}
\includegraphics[scale=0.55]{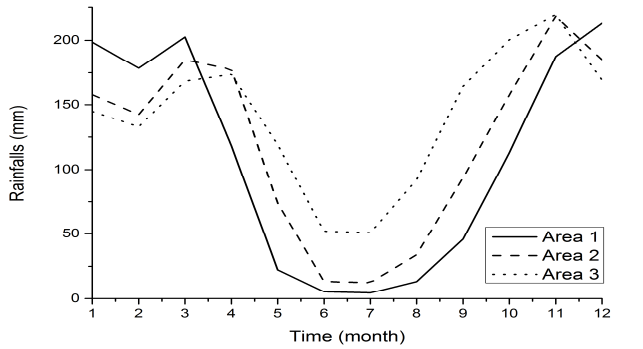}
\par\end{centering}
\caption{\label{fig:Typology-of-rainfall}Typology of rainfall regimes over
the Kasaï basin. Based on a hierarchical cluster analysis of mean
monthly rainfall for 1940-1999 (SIEREM.v1data).}
\end{figure}

\begin{figure}[H]
\begin{centering}
\includegraphics[scale=0.8]{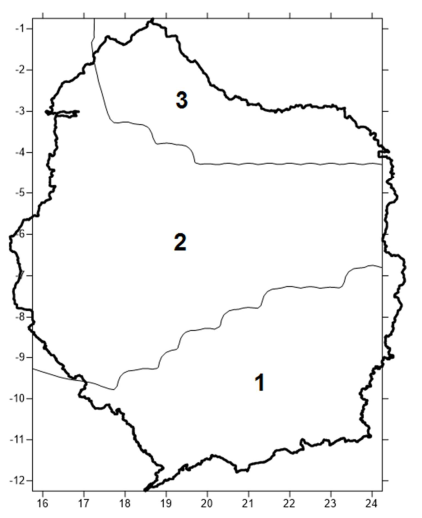}
\par\end{centering}
\caption{\label{fig:Map-of-the}Map of the 3 regional divisions of Kasaï basin
after hierarchical clustering methods on principal component analysis
of rainfall}
\end{figure}

\begin{figure}[H]
\begin{centering}
\includegraphics[scale=0.8]{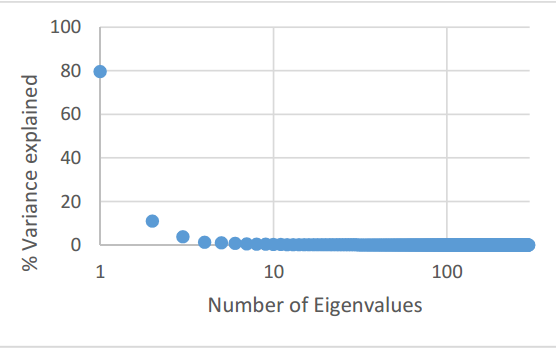}
\par\end{centering}
\caption{\label{fig:Percentage-of-variance}Percentage of variance explained}
\end{figure}

Only PC1 will be discussed as it contributes for 80 \% of the rainfall
variability in the Kasaï basin (figure \ref{fig:Percentage-of-variance}). 

\begin{figure}[H]
\begin{centering}
\includegraphics[scale=0.6]{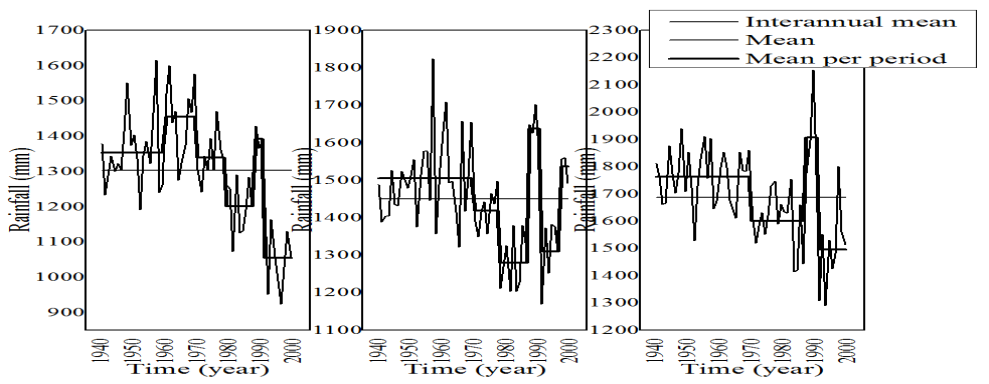}
\par\end{centering}
\caption{\label{fig:Phases-of-homogeneous}Phases of homogeneous rainfalls
over area 1,2 and 3 of Kasaï basin (1940-1999)}

\end{figure}

By Hubert segmentation, which gives us the year of different ruptures
observed on the rainfall time series (Fig. \ref{fig:Phases-of-homogeneous}),
we can see that rains on the Kasaï basin are marked by fluctuations
with a tendency on the one hand to the noticeable decline between
1970 and 1990, 1993 and 1999 and also an upward trend between 1960
and 1970 in area 1, 1990 and 1992.

The rain in area 1 presents five ruptures and therefore six times
homogeneous rainfall:

wet phases 1940-1959; 1959-1969; 1970-1979; 1988-1990; dry phases
1980-1987; 1991-1999;

The rain in area 2 presents five ruptures and therefore six times
homogeneous rainfall:

wet phases 1940-1969; 1988-1990; 1996-1999; dry phases 1970-1975;
1976-1987; 1991-1995;

The rain in area 3 presents three ruptures and therefore four times
homogeneous rainfall:

wet phases 1940-1959; 1970-1979; dry phases 1959-1969; 1980-1987;

\subsection{IDENTIFICATION OF VARIABILITY MODES}

The wavelet analysis is used to study the frequency composition and
visualize unsteady fluctuations of hydro variables so as to identify
the main modes of variability of flows and precipitation in the basin
of the Kasaï.

\begin{figure}[H]
\begin{centering}
\includegraphics[scale=0.8]{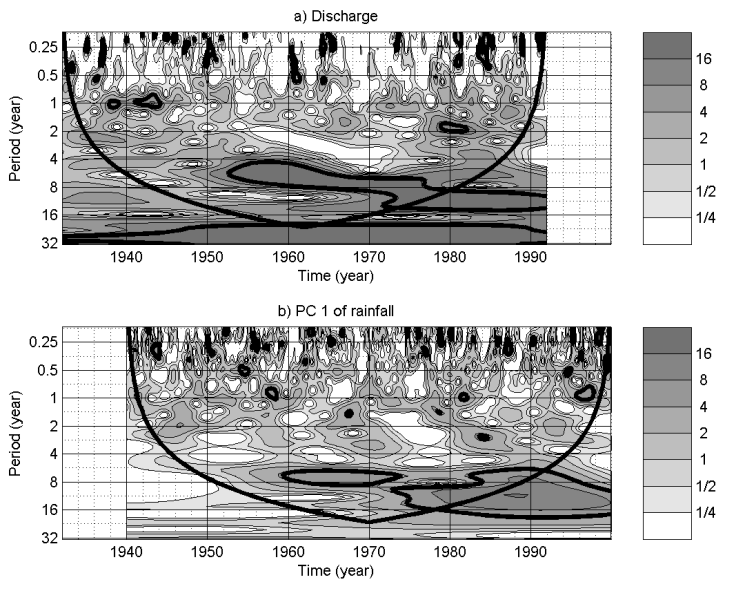}
\par\end{centering}
\caption{\label{fig:Continuous-wavelet-power}Continuous wavelet power spectrum
of a) discharge of Kasaï River at Ilebo station and b) PC 1 of gridded
rainfall over Kasaï basin. The left axis is the Fourier period (in
yr) corresponding to the wavelet scale on the right axis. The bottom
axis is time (yr). The thick black contour designates the 5\% significance
level against red noise and the cone of influence (COI) where edge
effects might distort the picture is shown as a lighter shade.}
\end{figure}

\begin{figure}[H]
\begin{centering}
\includegraphics[scale=0.75]{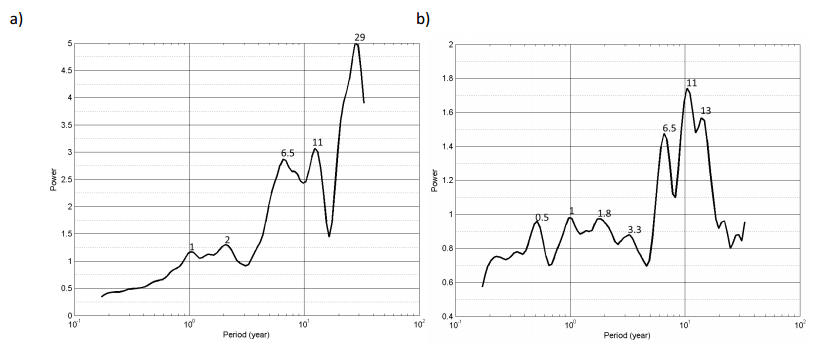}
\par\end{centering}
\caption{\label{fig:Map-of-the-1}Continuous global wavelet power spectrum
of a) discharge of Kasaï River at Ilebo station and b) PC 1 of gridded
rainfall over Kasaï basin.}

\end{figure}

Table 1 shows the contribution of the energy bands in the discharge
at Ilebo station and in rainfall in the Kasaï basin. The wavelet analysis
of Kasaï flow shows a plurality of energy bands (Figure \ref{fig:Continuous-wavelet-power}
and \ref{fig:Map-of-the-1}): A period of about 1 year mainly expressed
A band of 2-8 years well localized:
\begin{itemize}
\item between 1953 and 1973 for both rainfalls and discharge;
\item between 1973 and 1995 for rainfalls
\end{itemize}
A band of 8-16 years with a structure which begin around 1973 till
1999.

One can find a change point around 1970 (Figure \ref{fig:Continuous-wavelet-power}
and \ref{fig:Map-of-the-1}) which is associated to low rainfalls
and discharge. 

\begin{table}[H]

\caption{\label{tab:Contribution-of-selected}Contribution of selected energy
bands on total variance of discharges for the Kasaï River at Ilebo
station and rainfalls in Kasaï basin}

\begin{centering}
\begin{tabular}{|c|c|c|c|c|}
\hline 
 & Discharge & \multicolumn{3}{c|}{Rainfall}\tabularnewline
\hline 
\hline 
 & Ilebo & Region 1 & Region 2 & Region 3\tabularnewline
\hline 
0-1 year & 42\% & 44\% & 45\% & 45\%\tabularnewline
\hline 
1-2 years & 34\% & 35\% & 30\% & 25\%\tabularnewline
\hline 
2-8 years & 10\% & 5\% & 6\% & 8\%\tabularnewline
\hline 
8-16 years & 6\% & 3\% & 4\% & 6\%\tabularnewline
\hline 
\end{tabular}
\par\end{centering}
\end{table}

\subsection{COHERENCE BETWEEN RAINFALLS AND DISCHARGES}

\begin{figure}[H]
\begin{centering}
\includegraphics[scale=0.6]{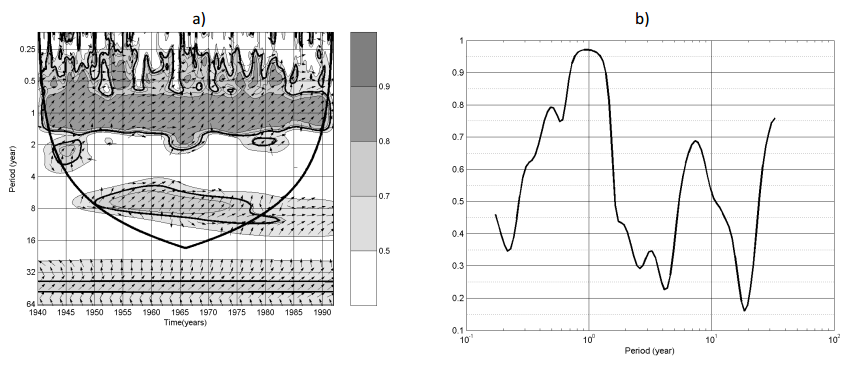}
\par\end{centering}
\caption{\label{fig:Wavelet-coherence-between}Wavelet coherence between the
discharge of Kasaï River at Ilebo station and the PC 1 (80\%) of rainfalls
over Kasaï basin. The 5\% significance level against red noise is
shown as a thick contour. The relative phase relationship is shown
as arrows (with in-phase pointing right, anti-phase pointing lefting
straight down).}

\end{figure}

The discharge of Kasaï River at Ilebo station seems to be strongly
related to the rainfalls with total coherence around 59.63\%. Figure
\ref{fig:Wavelet-coherence-between} attests that the coherence between
discharge and rainfall in the Kasaï basin are essentially distributed
across one year cycle and in 2-8 band (6.5 years). 

\subsection{RELATIONSHIP TO CLIMATE FORCING}

\begin{figure}[H]
\begin{centering}
\includegraphics[scale=0.8]{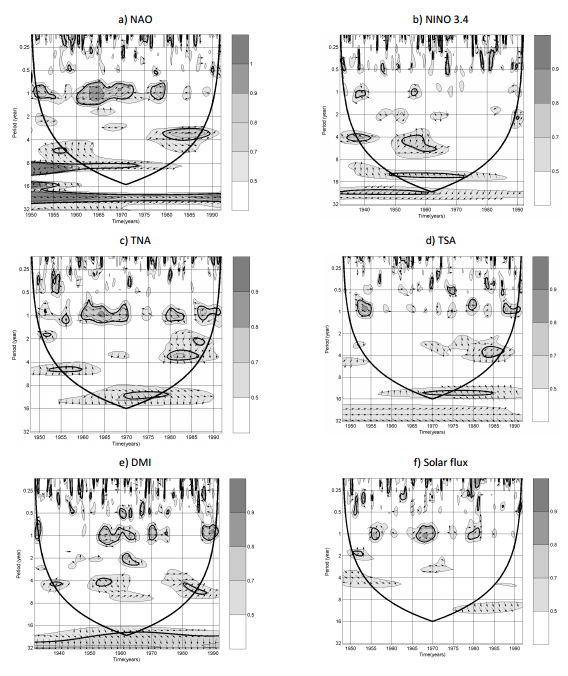}
\par\end{centering}
\caption{\label{fig:Wavelet-coherence-between-1}Wavelet coherence between
the discharge of Kasaï River at Ilebo station and climate indices
(NAO, Nino 3.4, TNA, TSA, DMI, Solar flux. The 5\% significance level
against red noise is shown as a thick contour.}

\end{figure}

\begin{figure}[H]
\begin{centering}
\includegraphics{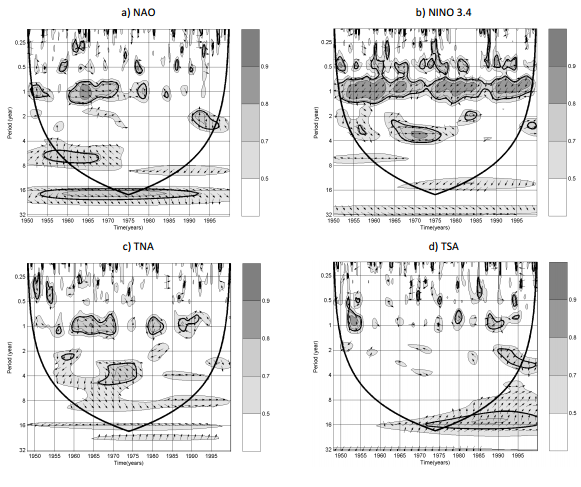}
\par\end{centering}
\begin{centering}
\includegraphics{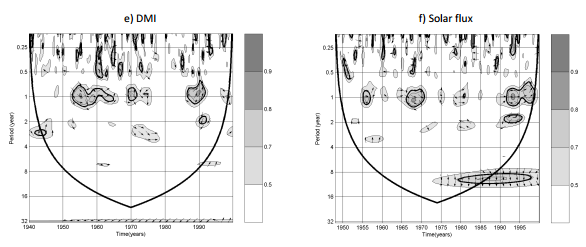}
\par\end{centering}
\caption{\label{fig:Wavelet-coherence-between-2}Wavelet coherence between
the PC 1 of rainfalls over Kasaï basin and climate indices (NAO, Nino
3.4, TNA, TSA, DMI, Solar flux. The 5\% significance level against
red noise is shown as a thick contour. The relative phase relationship
is shown as arrows (with in-phase pointing right, anti-phase pointing
left, and BMI leading AO by 90\textopenbullet{} pointing straight
down).}

\end{figure}

Some characteristic oscillations were next related to several well-known
climate indices. The coherence wavelet analyses between discharge
time series at Ilebo station, rainfalls (Figure \ref{fig:Wavelet-coherence-between})
in the three area of Kasaï basin and climate index (NAO, NINO 3.4,
TNA, TSA, DMI and Solar flux) show period of high coherence during
some periods listed in table \ref{tab:-Contribution-of} and \ref{tab:Contribution-of-selected-1}.
Figure \ref{fig:Wavelet-coherence-between-1} and \ref{fig:Wavelet-coherence-between-2}
also shows the characteristic periods of high wavelet coherence between
discharge at Ilebo station, first principal component of rainfalls
in Kasaï basin and selected climate indices. 

\begin{table}[H]
\caption{ \label{tab:-Contribution-of}Contribution of selected energy bands
on total variance of discharges for the Kasaï River at Ilebo station
and rainfalls in Kasaï basin}

\begin{centering}
\begin{tabular}{|c|c|c|c|c|c|c|}
\hline 
Observed scales & NAO & Nino 3.4 & TNA & TSA & DMI & Solar\tabularnewline
\hline 
\hline 
2-4 years & 1951-1955 &  & 1982-1987 & 1982-1987 & 1960-1969 & \tabularnewline
\hline 
 & 1965-1971 &  &  &  &  & \tabularnewline
\hline 
 & 1979-1990 &  &  &  &  & \tabularnewline
\hline 
4-8 years & 1955-1958 & 1932-1945 &  &  & 1950-1957 & 1948-1962\tabularnewline
\hline 
 &  & 1952-1962 &  &  & 1980-1990 & \tabularnewline
\hline 
8-16 years & 1950-1960 & 1948-1972 & 1967-1985 & 1967-1985 &  & \tabularnewline
\hline 
\end{tabular}
\par\end{centering}
\end{table}

\begin{table}[H]
\caption{\label{tab:Contribution-of-selected-1}Contribution of selected energy
bands on total variance of discharges for the Kasaï River at Ilebo
station and rainfalls in Kasaï basin }

\centering{}%
\begin{tabular}{|c|c|c|c|c|c|c|}
\hline 
{\small{}Observed scales} & {\small{}NAO} & {\small{}Nino 3.4} & {\small{}TNA} & {\small{}TSA} & {\small{}DMI} & {\small{}Solar flux}\tabularnewline
\hline 
\hline 
\multirow{4}{*}{{\small{}2-4 years}} &  & {\small{}1955-1957} & {\small{}1957-1960 } &  & {\small{}1940-1948 } & {\small{}1956-1961}\tabularnewline
\cline{2-7} 
 &  & {\small{} 1964-1975} & {\small{}1967-1976} &  & {\small{}1951-1952 } & \tabularnewline
\cline{2-7} 
 &  &  & {\small{}1979-1984} &  & {\small{}1970-1978} & \tabularnewline
\cline{2-7} 
 &  &  & {\small{}1982-1992} &  &  & \tabularnewline
\hline 
{\small{}4-8 years} & {\small{}1954-1967} & {\small{}1950-1967} & {\small{}1962-1972} &  &  & \tabularnewline
\hline 
{\small{}8-16 years} & {\small{}1975-1995} & {\small{}1967-1999} & {\small{}1957-1999} & {\small{}1972-1999} &  & {\small{}1980-1997}\tabularnewline
\hline 
\end{tabular}
\end{table}

\begin{figure}[H]
\begin{centering}
\includegraphics{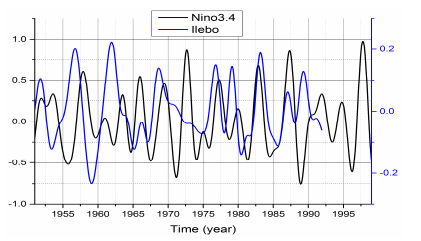}
\par\end{centering}
\caption{\label{fig:Comparison-between-the}Comparison between the wavelet
\textendash filtered reconstructions of the monthly discharge over
2-8 band-passed of Kasaï river at Ilebo station and Nino 3.4 index.}

\end{figure}

\begin{figure}[H]
\begin{centering}
\includegraphics{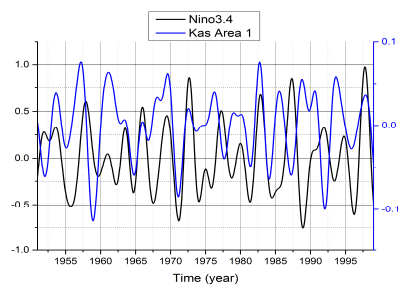}
\par\end{centering}
\caption{\label{fig:Comparison-between-the-1}Comparison between the wavelet
\textendash filtered reconstructions of the monthly rainfall over
2-8 band-passed on area 1 of Kasaï basin and Nino 3.4 index. }

\end{figure}

\begin{figure}[H]
\begin{centering}
\includegraphics{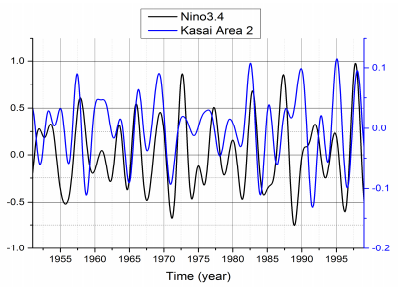}
\par\end{centering}
\caption{\label{fig:Comparison-between-the-2}Comparison between the wavelet
\textendash filtered reconstructions of the monthly rainfall over
2-8 band-passed on area 2 of Kasaï basin and Nino 3.4 index. }

\end{figure}

\begin{figure}[H]
\begin{centering}
\includegraphics[scale=0.6]{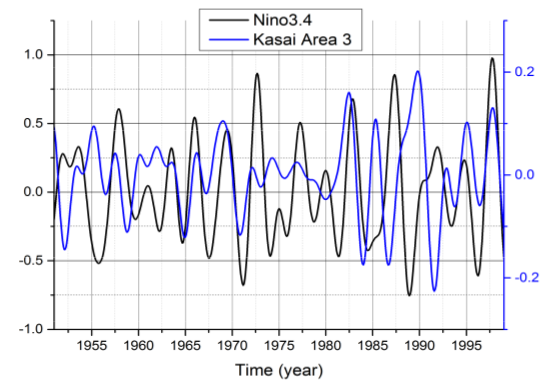}
\par\end{centering}
\caption{\label{fig:Comparison-between-the-3}Comparison between the wavelet
\textendash filtered reconstructions of the monthly rainfalls over
2-8 band-passed on area 3 of Kasaï basin and Nino 3.4 index.}

\end{figure}

\begin{figure}[H]
\begin{centering}
\includegraphics{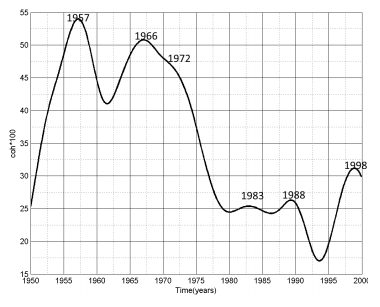}
\par\end{centering}
\caption{\label{fig:Percentage-of-scale-averaged}Percentage of scale-averaged
wavelet coherence over the 2\textendash 8-yr band for the gridded
rainfall over Kasaï basin and the Niño 3.4 SST index.}

\end{figure}

Figure \ref{fig:Percentage-of-scale-averaged} presents the relationships
over the 2\textendash 8-yr band between the first principle component
of the gridded rainfall over Kasaï basin and the Niño3.4 SST index. 

Mean percentage of wavelet coherence between discharge at Ilebo station
or rainfall and some selected climate indices except Nino 3.4 index
shows a weak nonlinear relationship (figure 20). 

\begin{figure}[H]
\begin{centering}
\includegraphics[scale=0.8]{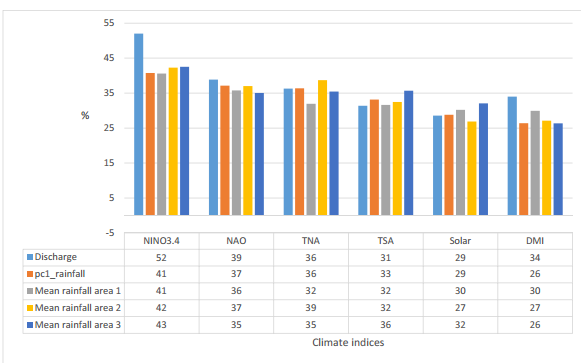}
\par\end{centering}
\caption{\label{fig:Mean-percentage-of}Mean percentage of wavelet coherence
( 2-8 years band-passed) between of Ilebo\textquoteright s discharge,
PC 1 of gridded rainfalls, Monthly mean rainfalls in Kasaï\textquoteright s
area 1, area 2, area 3 and some climate indices (NAO, NINO 3.4, TNA,
TSA, DMI and Solar flux)}

\end{figure}

\section{5 DISCUSSIONS AND CONCLUSION}

We propose in this article a time-scale analysis of the river discharge
fluctuations at Ilebo and rainfalls over the Kasaï basin. The two
temporal discontinuities around 1970-1975 and 1990-1995 over figure
\ref{fig:Phases-of-homogeneous} and \ref{fig:Continuous-wavelet-power}
were also found in other studies \citep{key-8}, \citep{key-19}.
The wavelet coherence between discharges at Ilebo and rainfall over
Kasaï River shows a high correlation in annual and the 2-8 year band
\textendash passed. Wavelet coherence between discharges at Ilebo
and rainfall over Kasaï River and selected climate index confirm a
less (figure \ref{fig:Mean-percentage-of}) and nonlinear relationship
between the hydrology of the Kasaï basin and the Pacific, Atlantic
and Indian SST fluctuations. One can see covariability between rainfalls
in the Kasaï basin and ENSO phenomenon, particularly at El Niño years
\citep{key-10} (Figure \ref{fig:Comparison-between-the}-\ref{fig:Percentage-of-scale-averaged}).
The influence of teleconnections is low but their effects seem especially
important having regard to the covariability observed with El Niño
example. Further investigations should be conducted to understand
the variability of the Kasaï river basin regime variability. What
is the role of deforestation and of human activity and urbanization
on water resource to availability?

\section*{ACKNOWLEDGMENTS}

Thanks for the gridded rainfall data providers IRD and HydroSciences
Montpellier, France. The wavelet software was provided by:

Torrence and Compo at http://atoc.colorado.edu/research/wavelets/ 

A. Grinsted at http://www.pol.ac.uk/home/research/waveletcoherence/

\end{document}